\documentclass[useAMS,usenatbib]{mn2e}
\usepackage{graphicx}
\usepackage[usenames,dvipsnames]{xcolor}
\usepackage{url}
\usepackage{amssymb}
\usepackage{lscape}

\newcommand{\kms}{~km s$^{-1}$~}
\newcommand{\WR}{WR~137~}
\newcommand{\WRE}{WR~137}
\newcommand{\dotM}{~M$_{\odot}$~yr$^{-1}$~}
\newcommand{\XMM}{{\it XMM-Newton~}}
\newcommand{\XMME}{{\it XMM-Newton}}
\newcommand{\Chandra}{{\it Chandra~}}

\newcommand{\apj}{ApJ}
\newcommand{\xspec}{{\sc xspec~}}
\newcommand{\xspecE}{{\sc xspec}}

\def\kms{\mbox{~km\,s$^{-1}$\/}}

\def\utw{\smash{\rlap{\lower5pt\hbox{$\sim$}}}}
\def\udtw{\smash{\rlap{\lower6pt\hbox{$\approx$}}}}

\title[X-rays from \WR]
{X-rays from the episodic dust maker 
\WR }
\author[S.A.Zhekov]{Svetozar A. Zhekov\thanks{E-mail: 
szhekov@space.bas.bg} \\
Space Research and Technology Institute, Akad. G.
Bonchev str., bl.1, Sofia 1113, Bulgaria\\
}

\date{}

\pagerange{\pageref{firstpage}--\pageref{lastpage}} \pubyear{2002}

\begin{document}

\maketitle

\label{firstpage}

\begin{abstract}
We present an analysis of the \XMM observation of the episodic dust
maker \WRE. Global spectral fits show that its X-ray spectrum 
is well matched by a two-temperature optically-thin 
plasma emission (kT$_1 \sim 0.4$ keV and kT$_2 \sim 2.2$ keV). 
If we adopt the colliding stellar wind (CSW) picture for this wide 
WR$+$O binary, the theoretical CSW spectra match well the shape of the
observed X-ray spectrum of \WR but they overestimate the observed flux 
(emission measure) by about two orders of magnitude.
To reconcile the model predictions with observations,
the mass loss of \WR must be reduced considerably (by about an order
of magnitude) with respect to its currently accepted value. In all
the spectral fits, the derived X-ray absorption is consistent with the
optical extinction to \WRE.

\end{abstract}


\section{Introduction}
\label{sec:intro}
Wolf-Rayet (WR) stars are  massive stars that are descendants from the
massive O-type stars. WR stars have powerful winds and are losing mass 
at high rates ($\dot{M}$ $\sim$ 10$^{-5}$ M$_{\odot}$ yr$^{-1}$;
V$_{wind} = 1000-3000$\kms). There are three spectral sequences
depending on the abundance of light metals:
nitrogen-rich (WN),  carbon-rich (WC) and oxygen-rich (WO) Wolf-Rayet
stars. The observed binary fraction in the WR stars in the Galaxy is
relatively high and about 40\% of them are members of WR$+$O systems
\citep{vdh_01}.

There is a class of WRs, so called episodic dust makers (EDM), which
show recurrent infrared bursts that fade away with time (e.g.
\citealt{williams_95}; 2008 and the references therein).
Seven WC stars are known EDMs. These are 
WR 19, WR 48a, WR 70, WR 98a, WR 125, WR 137 and WR 140.
All of these objects possess variable NIR/IR emission characterized by
at least one `burst', that is a rapid rise of their emission that
faded away in a period of time (a few months and even years). 
For some of these objects, it is
certain that the WC star is a member of a binary system with the same
period of their recurrent IR emission: WR 140 (\citealt{williams_90};
\citealt{williams_11} and the references therein), WR 137 
\citep{lefevre_05}. For others, binarity is suspected based on
presence of absorption lines in their optical spectra that are typical
for massive O stars or it is suggested by the dilution of their
spectral lines.

It is believed that colliding stellar winds (CSW) in massive WR
binaries play a key role for the physics of EDMs and more specifically
for the origin of their infrared, radio and X-ray emission. 
Apart from their periodic IR emission, the EDMs should posses strong
X-ray emission and they should be non-thermal radio
sources. These basic characteristics are best illustrated by the
prototype EDM object, the WR$+$O binary WR 140. They are reasonably
well explained as a result from CSWs in a wide binary system
with highly elliptical orbit \citep{williams_90}.

We recall that a very basic feature of CSWs is their strong X-ray
emission that originates from the interaction region of the
winds of the massive binary components
(\citealt{pri_us_76}; \citealt{cherep_76}).
This likely explains the fact that came out from 
the first systematic survey of WRs: the WR$+$O binaries 
are the brightest X-ray sources amongst WR stars \citep{po_87}.
In the framework of the CSW picture, one can thus expect that all the
episodic dust makers should possess an enhanced X-ray emission. 

The X-ray emission from two episodic dust makers is studied in
considerable detail.
The prototype EDM, WR 140, has been detected with almost all the X-ray
observatories (e.g., Williams et al. 1990; Zhekov \& Skinner 2000; 
Pollock et al. 2005; Williams  2011). It has variable X-ray
emission, variable X-ray absorption, its X-ray plasma is hot 
(kT~$\geq 3$~keV). 
The analysis of recent \XMM and \Chandra data on 
WR 48a showed that this is the most X-ray luminous WR star in the 
Galaxy detected so far, after the black hole candidate Cyg X-3 and 
its dominant temperature component is kT~$\approx 3$~keV 
(\citealt{zhgsk_11}; 2014).
The luminous X-ray emission and the high plasma temperature can be
considered as a solid sign for presence of CSWs in these objects.
Similar findings for other EDMs are thus needed to check the
validity of the CSW paradigm for this class of objects.

In this paper, we report results from the \XMM observation of \WRE, the
third object amongst the episodic dust makers with an acceptably good
quality of its X-ray spectra. The paper is organized as follows.
We give some basic information about \WR
in Section~\ref{sec:wrstar}. In Section~\ref{sec:observations}, we
review the \XMM observation. In Section~\ref{sec:results}, we present
results from analysis of the X-ray properties of \WRE. In
Section~\ref{sec:discuss}, we discuss our results, and we present our
conclusions in Section~\ref{sec:conclusions}.

\section{The Wolf-Rayet Star \WR}
\label{sec:wrstar}
\WR (HD 192641) is one of the seven WR stars that originally formed 
the group of episodic dust makers \citep{williams_95}. It is a
spectroscopic binary (WC7pd$+$O9) with an orbital period of
$4765\pm50$ days at a distance of 2.38 kpc \citep{vdh_01}. 
The binary period was derived from
analysis of the infrared variability, that is from the consecutive
dust-formation episodes \citep{williams_01}. It was later confirmed
by spectroscopic studies in the optical domain: $4766\pm66$ days
\citep{lefevre_05}.
The optical extinction toward \WR is A$_v = 1.97$~mag
(\citealt{vdh_01}; A$_v = 1.11$ A$_{\mbox{V}}$) 
implying a foreground column density of
N$_H = (2.94-3.94)\times10^{21}$~cm$^{-2}$.
The range corresponds to the conversion that is used:
N$_H = (1.6-1.7)\times10^{21}$A$_{\mbox{V}}$~cm$^{-2}$
(\citealt{vuong_03}, \citealt{getman_05});
and 
N$_H = 2.22\times10^{21}$A$_{\mbox{V}}$~cm$^{-2}$
\citep{go_75}.
We adopt the stellar wind
parameters (velocity and mass loss) of
V$_{wind} = 1885$\kms and 
$\dot{M} = 3\times10^{-5}$\dotM \citep{nugis_98}.

In radio, no in detail studies are carried out so far. \WR was 
classified as a thermal radio source by \citet{ab_86} and it was also
listed as a non-thermal radio source in \citet{do_00} but that
classification was based only on a private communication.

In X-rays, a marginal detection was reported from the {\it Einstein}
survey of Wolf-Rayet stars, L$_X = 1.6\pm1.1\times10^{32}$~ergs s$^{-1}$
\citep{po_87}, and also from the {\it ROSAT} survey of Wolf-Rayet 
stars,  L$_X = 0.61\pm0.10\times10^{32}$~ergs s$^{-1}$ \citep{po_95}.
Both data sets had poor photon statistics ($\leq 50$ source counts).

\begin{figure}
\begin{center}
\centering\includegraphics[width=\columnwidth]{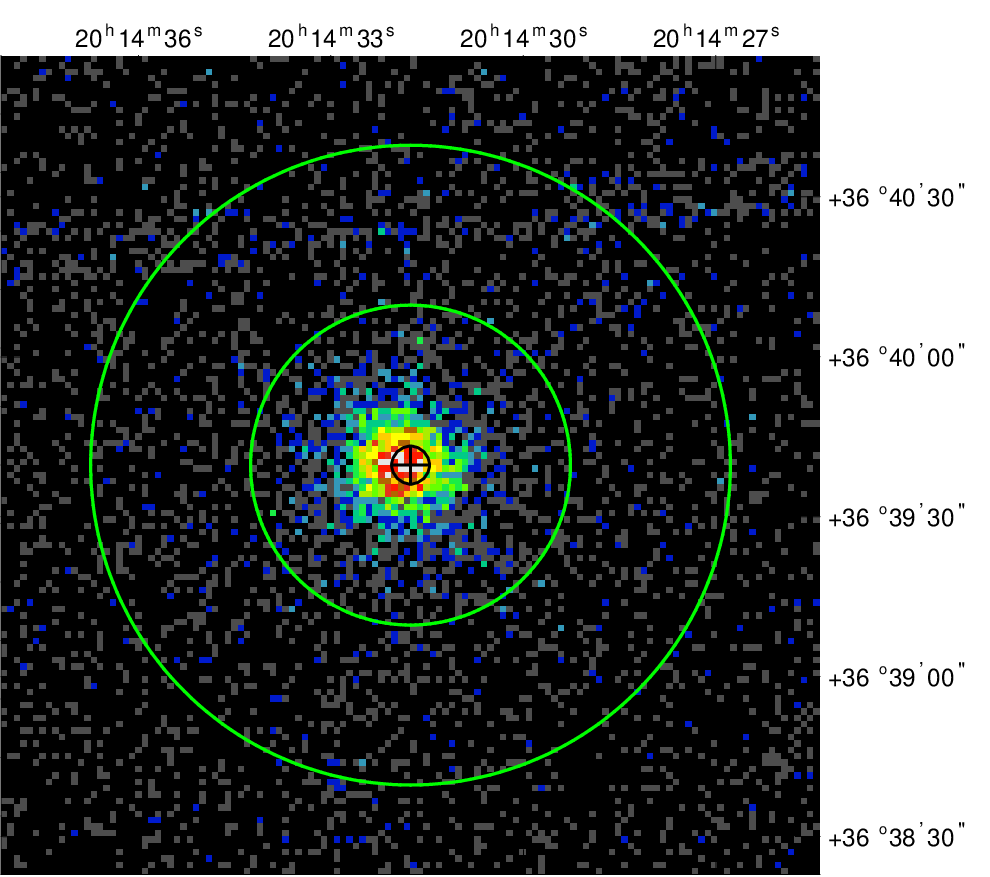}
\end{center}
\caption{The raw EPIC-pn image of \WR in the 0.2 - 10 keV energy
band with the spectral extraction regions. The source spectrum was
extracted from the central circle, while the background spectrum was
extracted from adjacent annulus. The circled plus sign gives the
optical position of \WR (SIMBAD).
}
\label{fig:image}
\end{figure}

\begin{figure}
\begin{center}
\centering\includegraphics[width=2.24in,
height=3.0in,angle=-90]{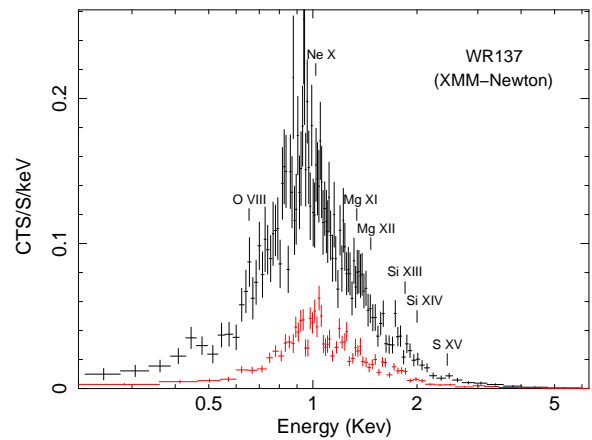}
\end{center}
\caption{The background-subtracted spectra of \WR re-binned to have a
minimum of 30 counts per bin. Positions of some, usually strong,
emission L$_{\alpha}$ and  K$_{\alpha}$ line features of various ionic
species are marked. 
The pn (upper curve) and MOS (lower curve) spectra are
shown in black and red colour, respectively.
}
\label{fig:spec_XMM}
\end{figure}

\section{Observations and data reduction}
\label{sec:observations}

\WR was observed with \XMM on 2013 Oct 9 (Observation ID 0720600101)  
with a nominal exposure of $\sim 50$ ks. The source was detected (see
Fig.~\ref{fig:image}) but it was not very bright in X-rays, thus, our 
analysis is based on the data from the European Photon Imaging Camera 
(EPIC) having one pn and two MOS detectors\footnote{see \S~3.3 in the 
\XMM Users Handbook,
http://xmm.\\esac.esa.int/external/xmm\_user\_support/documentation/uhb
}.
For the data reduction, we made use of the \XMM 
{\sc sas}\footnote{Science Analysis Software, 
http://xmm.esac.esa.int/sas} 12.0.1 data analysis software.
The {\sc sas} pipeline processing scripts emproc and epproc were
executed to incorporate the most recent calibration files (as of 2014
June 4). The data were then filtered for high X-ray background 
following the instructions in the {\sc sas} documentation. The 
{\sc sas} procedures rmfgen and arfgen were adopted to generate the
corresponding response matrix files and ancillary response files for 
each spectrum. The MOS spectrum in our analysis is the sum of the
spectra from the two MOS detectors. The extracted spectra (0.2 - 10
keV) of \WR had
$\sim 3427 $~source counts in the 26.8-ks pn effective exposure and 
$\sim 2366$~source counts in the 32.6-ks MOS effective exposure.

Also, we constructed the pn and MOS1,2 background-subtracted light 
curves of \WRE. On a timescale less than 35 ks and time bins
between 100 and 1000 s, the X-ray light curves were statistically 
consistent with a constant flux:  
adopting $\chi^2$ fitting, the light curves
were fitted with a constant and the goodness-of-fit was $\geq 0.70$. 

Thus, our study was centred around the global analysis of the \XMM
spectra of \WRE.
For that, we used standard as well as custom
models in version 11.3.2 of \xspec \citep{Arnaud96}.

\section{Results}
\label{sec:results}
The X-ray spectra of \WR are shown in Fig.~\ref{fig:spec_XMM}. It is
worth noting that the emission line features are rather weak.
This is a bit in contrast with the situation in the X-ray spectrum of
another dust maker, WR 48a (see fig. 2 in \citealt{zhgsk_11}). We note
that the data for \WR and WR 48a were taken with exactly the same 
instrumentation (\XMM EPIC).

To derive the properties of the X-ray emitting plasma in \WRE, we
adopted global spectral fitting to its \XMM spectra. The EPIC-pn
and EPIC-MOS spectra were fitted simultaneously sharing identical
model parameters.
We used discrete-temperature models in \xspec that assume plasma in 
collisional ionization equilibrium (CIE; model {\it vapec}) as well as
such that assume plasma with non-equilibrium ionization (NEI; model
{\it vpshock}).
The adopted set of abundances is that from \citet{vdh_86} typical for
the WC stars (by number):
H~$=0.00$, He~$=0.618$, C~$=0.248$, N~$=0.00$, O~$=0.120$, 
Ne~$=1.15\times10^{-2}$, Mg~$=1.68\times10^{-3}$,
Si~$=4.23\times10^{-4}$, S~$=9.40\times10^{-5}$,
Fe~$=2.36\times10^{-4}$.
To improve the quality of the fits, the O, Ne, Mg, Si and Fe 
abundances were allowed to vary.
The emission plasma components were subject to common X-ray absorption
(model {\it wabs} in \xspecE).

\begin{figure*}
\begin{center}
\centering\includegraphics[width=2.24in,
height=3.0in,angle=-90]{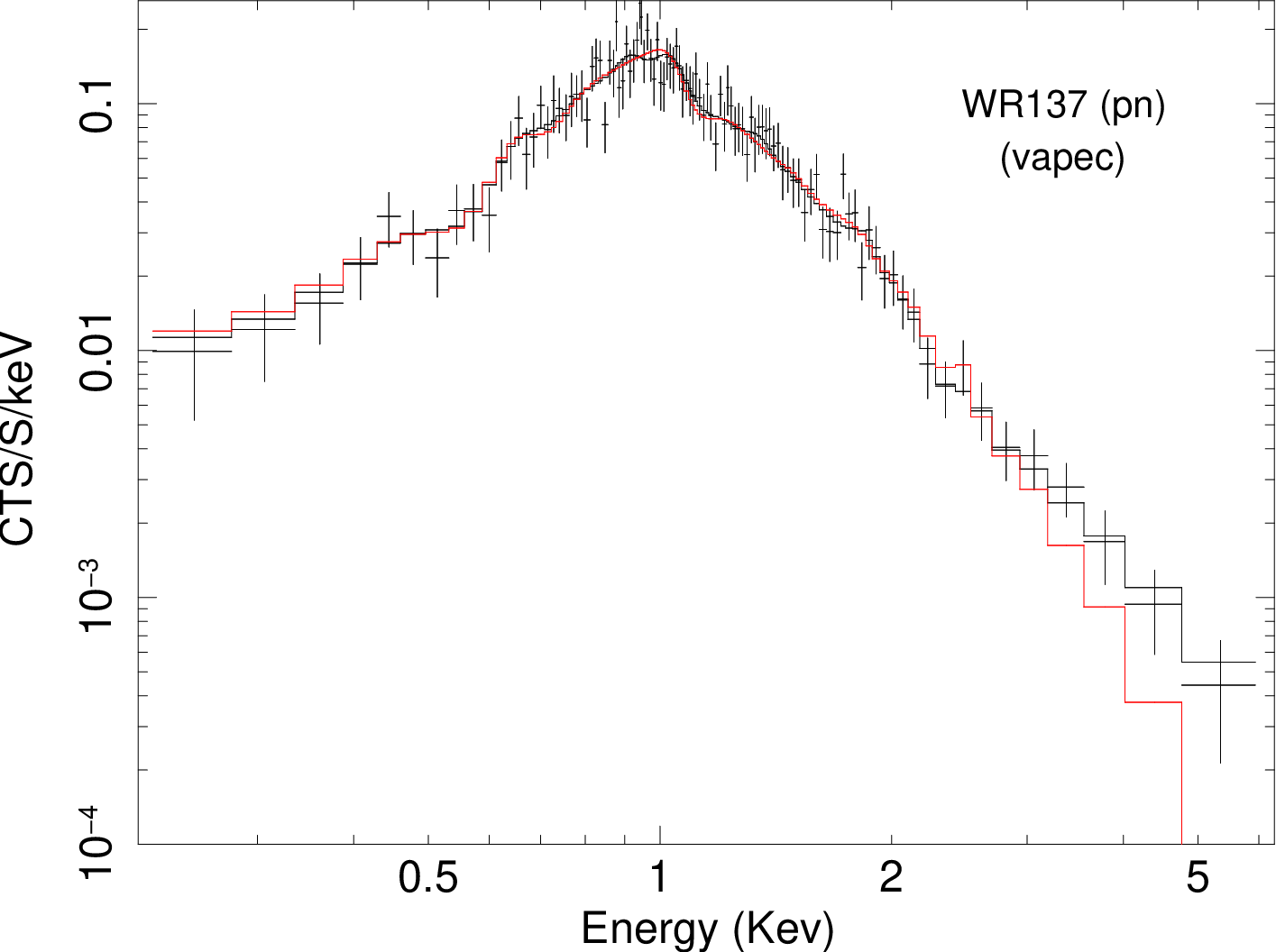}
\centering\includegraphics[width=2.24in,
height=3.0in,angle=-90]{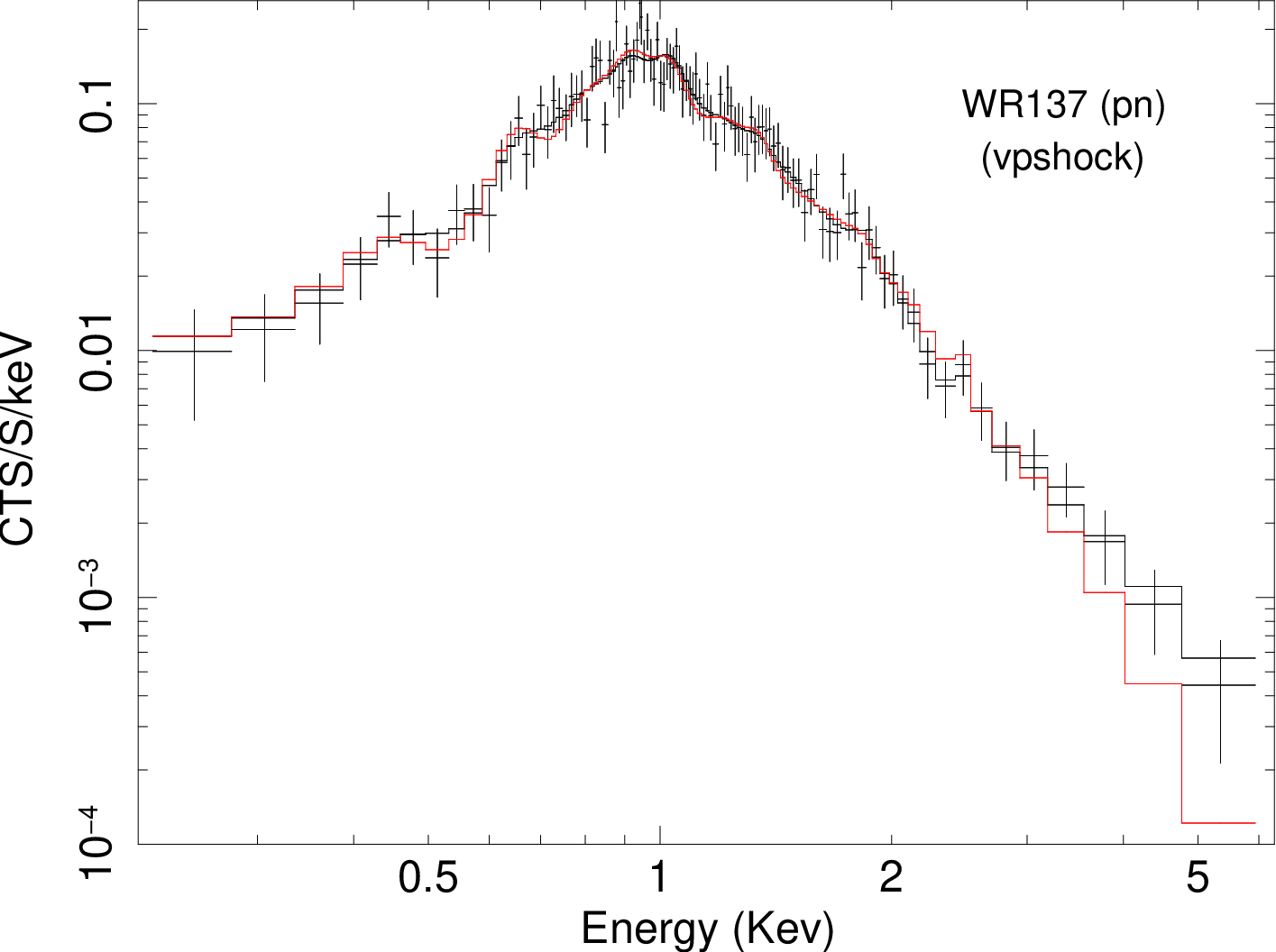}
\centering\includegraphics[width=2.24in,
height=3.0in,angle=-90]{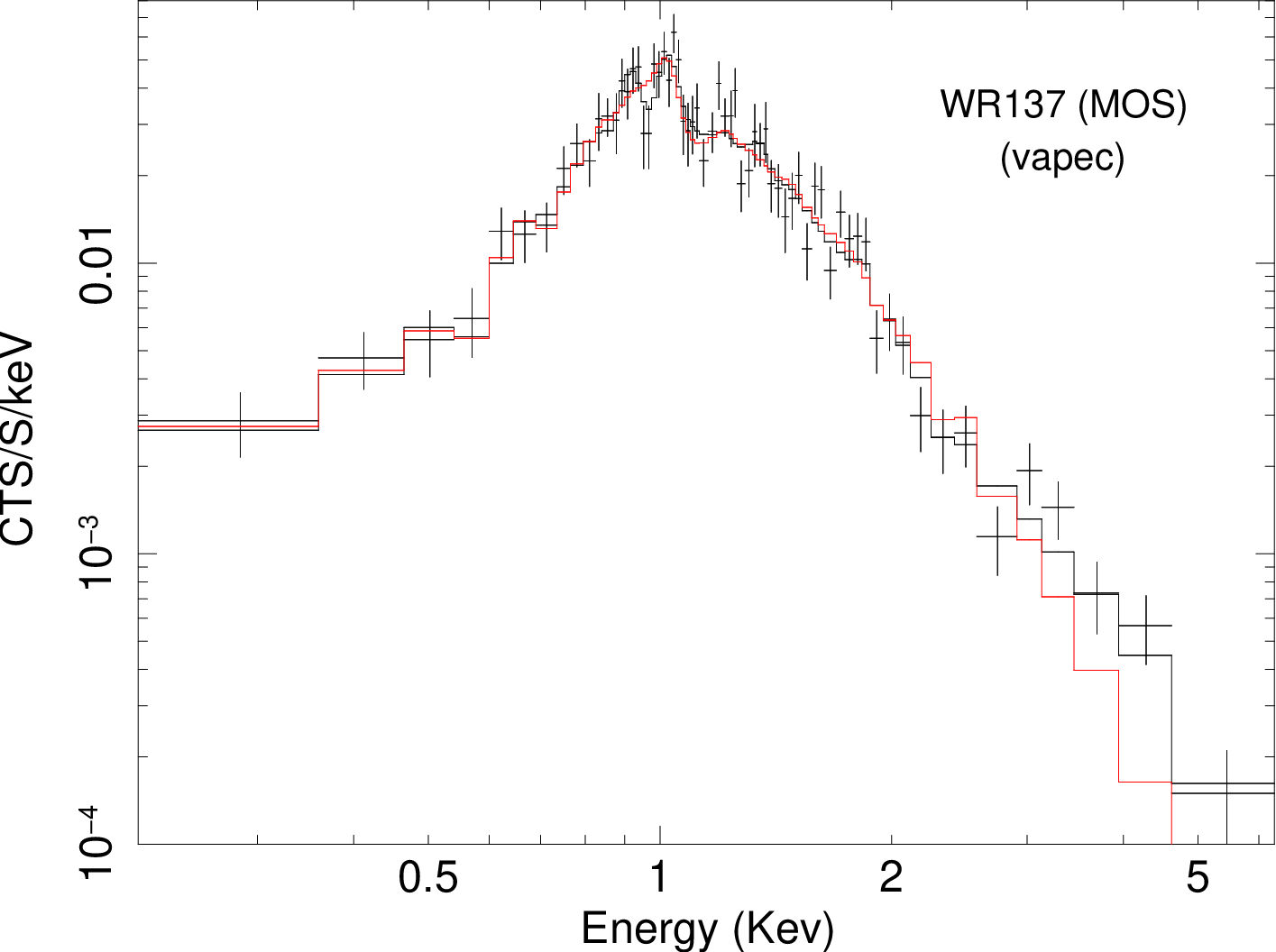}
\centering\includegraphics[width=2.24in,
height=3.0in,angle=-90]{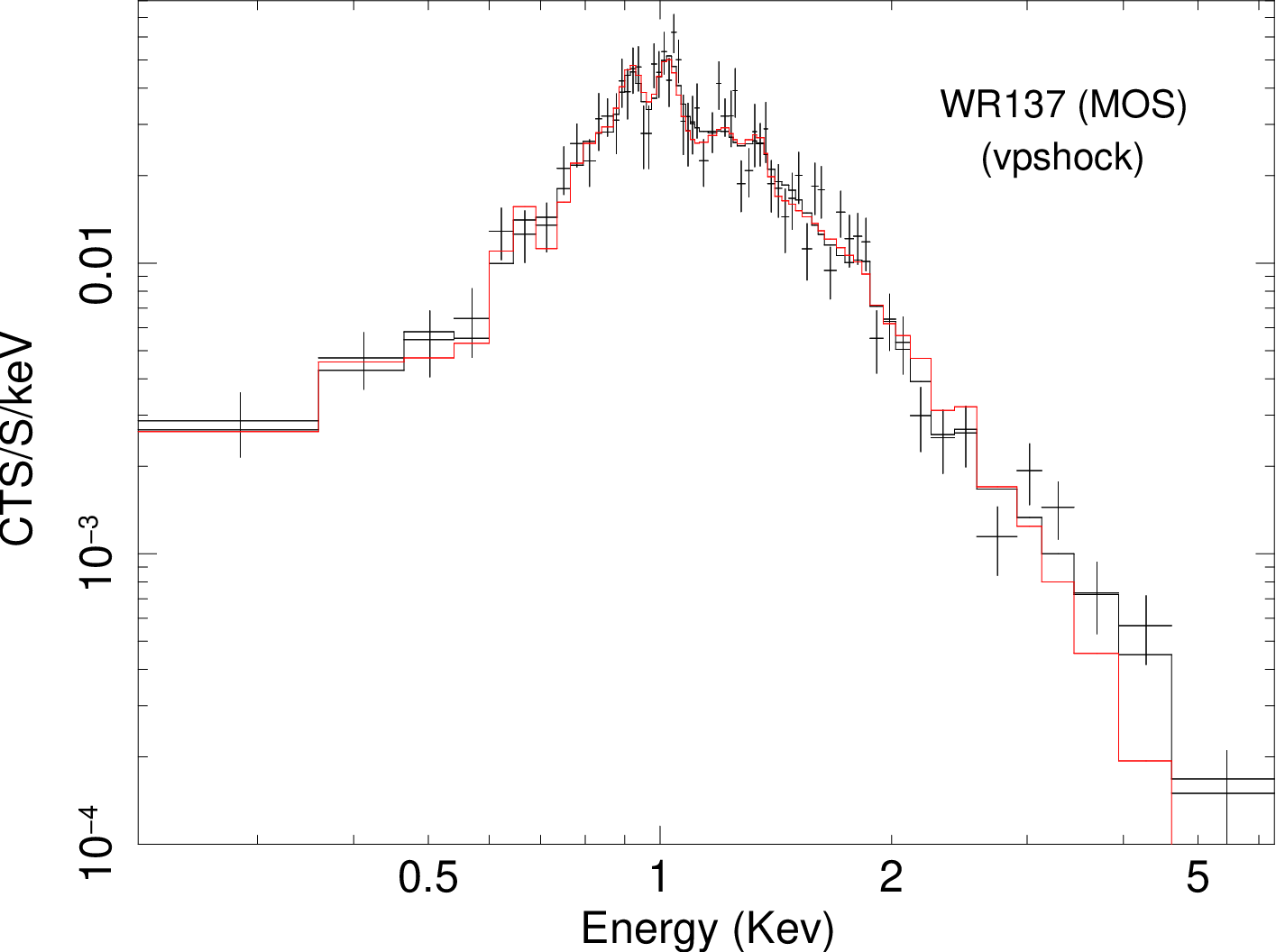}
\end{center}
\caption{The background-subtracted spectra of \WR and  the
discrete-temperature
model fits (Table~\ref{tab:fits}): `vapec' and 'vpshock' denote the
case of optically-thin plasma or plane-parallel shock emission,
respectively. In each panel, the two- and one-temperature model
spectra are shown with a stepped line in black and red colour,
respectively.
The spectra were re-binned to have a minimum of 30 counts per bin.
}
\label{fig:spec}
\end{figure*}

We performed one- and two-temperature plasma model fits. 
Table~\ref{tab:fits} and Figure~\ref{fig:spec} present 
the corresponding results from the fits to the \XMM spectra of \WRE.
A few things are worth mentioning.

We see that the one-temperature plasma models, with not very high
temperature (kT $\approx 0.8$~keV), give statistically acceptable fits 
to the X-ray spectra of \WR but they slightly underestimate its X-ray
emission at energies above 3 keV. On the other hand, the
spectral fits of the two-temperature plasma models do not have this 
caveat and the quality of the fits improves: the $\chi^2$ value 
decreases by 25-30\%. Thus, we will further focus mostly on the 
results from these model fits.

An interesting feature of the two-temperature NEI model fit is the 
relatively high value of the ionization age (model parameter $\tau$). 
We note that both emission components have such high values which is 
indicative of plasma with ionization state close to CIE. 

The X-ray absorption, as derived from the two-temperature fits, is in
general consistent with the optical extinction to \WR (see
Section~\ref{sec:wrstar}). Namely, it is only by 15-25\% higher than
the latter if we adopt the \citet{go_75} conversion. But the result may
indicate some extra absorption ($\sim 60$\%) if a more recent
conversion is used (\citealt{vuong_03}, \citealt{getman_05}; see
Section~\ref{sec:wrstar}). 

To explore this a bit in detail, we ran
the same two-temperature CIE and NEI models with two-component X-ray
absorption. The first component was fixed to the value corresponding
to the optical extinction (the ISM absorption component) and the
second component (the wind absorption component) was allowed to vary.
For physical consistency, the wind absorption component shared the
same abundances with the emission plasma components in the fits.
These fits were statistically as good as the ones with a single X-ray
absorption component (Table~\ref{tab:fits}) with $\chi^2/\mbox{dof} = 
119/162; ~118/160$ for the CIE and NEI models, respectively. The 
helium column density of the wind absorption was 
N$_{He, wind} = (0.5-1.0)\times10^{19}$ cm$^{-2}$. Such a low value of
the wind-absorption column density shows that the X-ray emission
region in \WR is {\it not} located deep in the WC wind. For comparison,
the radial helium column density of the WC wind from infinity to a
distance $R_{au}$ (given in au) is
N$_{He, wind, rad.} = 2.58\times10^{21} R^{-1}_{au}$ cm$^{-2}$,
if adopting the wind parameters of \WR (Section~\ref{sec:wrstar})
and the WC abundances. 
The column density derived from the fits then puts the X-ray
emission region at a very large (and unrealistic) distance of more
than 200 au from the WC star in \WRE. This is likely a sign that the
X-ray small extra (or wind) absorption is a result from the use of 
simple models in the spectral fits. Thus, it is conclusive that the 
X-ray absorption of \WR is consistent with the optical extinction 
towards this object.

We note that in general the X-ray spectrum of \WR does not show strong 
emission line features (Fig.~\ref{fig:spec_XMM}). As already noted, 
the second component in the two-temperature optically-thin plasma 
models mostly contributed to the relatively weak continuum at energies
higher than 3 keV. We ran two-component fits with an 
optically-thin plasma model ({\it vapec}
or {\it vpshock}) as a first component and a continuum model (a black
body  or a power-law model) as a second component. In statistical
sense, these fits proved to be equally successful as the 
two-temperature optically-thin plasma models.

The two-component fit, that is a sum of an optically-thin 
plasma and a black body model subject to common absorption, had 
$\chi^2/\mbox{dof} = 
119/162$, N$_{H} = 4.82^{+0.26}_{-0.21}\times10^{21}$ cm$^{-2}$
kT$_1 = 0.41^{+0.02}_{-0.02}$ keV, kT$_{BB} = 0.74^{+0.03}_{-0.02}$ keV, 
And, the corresponding fit with a power-law model as the second
component had $\chi^2/\mbox{dof} = 124/162$,  N$_{H} =
4.70^{+0.36}_{-0.25}\times10^{21}$ cm$^{-2}$, kT$_1 =
0.40^{+0.02}_{-0.03}$ keV, $\Gamma = 2.80^{+0.34}_{-0.17}$, where 
$\Gamma$ is the photon power-law index (F$_{PL} \propto E^{-[\Gamma-1]})$.
It is interesting to note that the values of the X-ray absorption and
the plasma temperature of the first component are practically
identical to the corresponding values from the fits with the 
two-temperature optically-thin plasma models (see Table~\ref{tab:fits}).

Finally, using the unabsorbed fluxes (two-temperature models in 
Table~\ref{tab:fits}) and a distance of 2.38 kpc 
(Section~\ref{sec:wrstar}) we see that the X-ray luminosity of \WR is 
not very high: $\log L_X = 32.76 - 32.83$ (L$_X$ in ergs s$^{-1}$).
In fact, its value is close to that typical for close WR$+$O binaries
(e.g., \citealt{zh_12}) and it is also similar to the upper values for
the range of X-ray luminosities amongst the presumably single WN stars
\citep{sk_10}. On the other hand, it is by one-two orders of
magnitude less than the X-ray luminosity of other dust makers as
WR140 (\citealt{zhsk_00}, \citealt{po_05}) and WR 48a
\citep{zhgsk_11}.

\begin{table*}
\caption{Global Spectral Model Results 
\label{tab:fits}}
\begin{tabular}{lllll}
\hline
\multicolumn{1}{c}{Parameter} & \multicolumn{1}{c}{2T vapec}  & 
\multicolumn{1}{c}{2T vpshock} &
\multicolumn{1}{c}{1T vapec}  & \multicolumn{1}{c}{1T vpshock} \\
\hline
$\chi^2$/dof  & 119/162 & 118/160 & 163/164 &  155/163  \\
N$_{H}$ (10$^{21}$ cm$^{-2}$)  & 
          4.66$^{+0.25}_{-0.16}$ & 5.15$^{+0.58}_{-0.43}$ & 
          3.51$^{+0.15}_{-0.15}$ & 6.30$^{+0.24}_{-0.35}$ \\ 
kT$_1$ (keV) & 0.40$^{+0.01}_{-0.02}$ & 0.37$^{+0.05}_{-0.02}$ & 
               0.79$^{+0.03}_{-0.03}$ & 0.82$^{+0.04}_{-0.03}$ \\ 
kT$_2$ (keV) & 2.11$^{+0.38}_{-0.23}$ & 2.24$^{+0.51}_{-0.42}$ & 
                                      &                    \\ 
EM$_1$ ($10^{54}$~cm$^{-3}$) &  2.09$^{+0.19}_{-0.10}$ &  
                                2.42$^{+0.53}_{-0.43}$ & 
                                1.57$^{+0.13}_{-0.12}$ & 
                                1.95$^{+0.18}_{-0.26}$ \\
EM$_2$ ($10^{54}$~cm$^{-3}$) &  0.51$^{+0.09}_{-0.03}$ & 
                                0.50$^{+0.10}_{-0.10}$ & 
                                                         & \\
$\tau_1$ ($10^{12}$ cm$^{-3}$ s)  &   &  8.62$^{+41.4}_{-5.54}$ & 
                                  &  0.21$^{+0.05}_{-0.04}$ \\ 
$\tau_2$ ($10^{12}$ cm$^{-3}$ s)  &   &  0.91$^{+49.1}_{-0.58}$ & 
                                  & \\ 
O   & 0.03$^{+0.02}_{-0.01}$ & 0.03$^{+0.01}_{-0.01}$  
    & 0.08$^{+0.03}_{-0.03}$ & 0.05$^{+0.01}_{-0.01}$  \\ 
Ne  & 0.12$^{+0.02}_{-0.02}$ & 0.10$^{+0.03}_{-0.02}$  
    & 0.11$^{+0.03}_{-0.03}$ & 0.07$^{+0.01}_{-0.01}$  \\ 
Mg  & 0.07$^{+0.04}_{-0.04}$ & 0.06$^{+0.02}_{-0.03}$  
    & 0.02$^{+0.03}_{-0.02}$ & 0.04$^{+0.01}_{-0.01}$  \\ 
Si  & 0.48$^{+0.24}_{-0.20}$ & 0.34$^{+0.28}_{-0.14}$  
    & 0.06$^{+0.08}_{-0.06}$ & 0.05$^{+0.05}_{-0.05}$  \\ 
Fe  & 0.23$^{+0.09}_{-0.08}$ & 0.26$^{+0.13}_{-0.10}$  
    & 0.36$^{+0.06}_{-0.05}$ & 0.49$^{+0.11}_{-0.20}$  \\ 
F$_{X}$ ($10^{-13}$ ergs cm$^{-2}$ s$^{-1}$)  & 
           2.61 (8.37) & 2.62 (10.0) & 
           2.38 (5.42) & 2.41 (16.0) \\ 
F$_{X,hot}$ ($10^{-13}$ ergs cm$^{-2}$ s$^{-1}$)  & 
           1.01 (1.58) & 1.10 (2.03) & 
                       &            \\ 
\hline

\end{tabular}

Note --
Results from  simultaneous fits to the EPIC 
spectra of \WRE.
Tabulated quantities are the neutral hydrogen absorption column
density (N$_{H}$), plasma temperature (kT), 
emission measure ($\mbox{EM} = \int n_e n_{He} dV $)
for a reference distance of d~$=2.38$~ kpc, 
shock ionization age
($\tau = n_e t$), the absorbed X-ray flux (F$_X$) in the 
0.5 - 10 keV range followed in parentheses by the unabsorbed value
(F$_{X,hot}$~ denotes the higher-temperature component, kT$_2$).
The derived abundances are with
respect to the typical WC abundances \citep{vdh_86}.
Errors are the $1\sigma$ values from the fits.

\end{table*}

\section{Discussion}
\label{sec:discuss}
Some of the basic results from the analysis of the X-ray
spectra of the dust maker \WR are as follows. 
The X-ray absorption is in general consistent with the optical
extinction to \WRE. The relatively hot plasma is present in the X-ray
emitting region but its emission does not dominate the X-ray
spectrum of this object. The X-ray luminosity is not very high: it
is considerably lower than that of dust makers with well studied X-ray 
emission (WR 48a and WR 140).

It is worth recalling that CSWs in wide WR$+$O binaries play an
important role for the physics of episodic dust makers. For example,
the strong and variable infrared, X-ray and non-thermal radio
emission of the prototype EDM, WR 140, can find an explanation 
in the CSW paradigm \citep{williams_90}. Studies of another EDM, WR
48a, suggest that CSWs should play an important role for its infrared
and X-ray emission (\citealt{williams_12}; \citealt{zhgsk_11}; 2014)
although the physical picture may be more complex
than that (see \citealt{zh_etal_14}). It is thus important to explore
the CSW picture for \WR in some detail as well.

\subsection{CSW model spectra}
\label{subsec:csw_spec}
To model the X-ray spectra from CSWs in \WRE, we used the models by
\citet{zhsk_00} and \citet{zh_07} that allow for different electron and 
ion temperatures and  the NEI effects. These models are based on the 
hydrodynamic model of adiabatic CSWs by \citet{mzh_93}. The latter 
assumes spherical symmetry of the stellar winds that have reached their 
terminal velocity before they collide. It also allows for different 
chemical abundances in the CSW region occupied by the shocked WR and 
O stellar winds, respectively. Thus, in all the spectral fits the
chemical abundances of the shocked O-star wind were solar \citep{an_89}
while those of the shocked WR-star wind were typical for the WC stars
\citep{vdh_86}. The latter were also allowed to vary in our fits to
the X-ray spectra of \WR in \xspecE.

We note that the basic input parameters for the CSW hydrodynamic model
in WR$+$O binaries are the mass loss and velocity of the stellar winds 
of the binary components and the binary separation. The former define a
dimensionless parameter 
$\Lambda = (\dot{M}_{WR} V_{WR}) / (\dot{M}_{O} V_{O})$ which
determines the shape and the structure of the CSW interaction region 
\citep{mzh_93}. 
 
For the wind parameters of the WC star in \WRE, we adopted 
V$_{WR} = 1885$\kms and $\dot{M}_{WR} = 3\times10^{-5}$ \dotM 
(Section~{\ref{sec:wrstar}). Using the orbital elements and
inclination angle of $i = 67^{\circ}$ (see \citealt{lefevre_05} and
table 4 therein) and Kepler's third law, we derived a mean binary 
separation of $a = 2.41\times10^{14}$~cm. We note that the time of the
\XMM observation (Section~\ref{sec:observations}) corresponds to an
orbital phase of $\phi = 0.338$ which 
for an orbital ellipticity of $e = 0.178$ 
translates into a binary separation of $1.113\times a$.
Since the wind parameters of
the O star in \WR are not observationally constrained, we adopted the
following approach in our procedure for modelling the CSW spectra of
this object. Namely, we assumed that the terminal wind velocity of the
O star is equal to that of the WC star in the system (V$_O =$ V$_{WR} =
1885$\kms). We explored a range of values for the
$\Lambda$-parameter: $\Lambda = 16,\, 25,\, 36$ (these correspond to
$\dot{M}_{O} = 1.88,\, 1.20,\, 0.83 \times10^{-6}$\dotM, respectively).
It is worth noting that the exact values of the O-star wind parameters
are not quite important for the X-ray emission from CSWs in WR$+$O
binaries because of the dominant contribution of the shocked
WR-wind gas \citep{mzh_93}. Anticipating the results from the CSW
spectral fits, we note that the contribution from the shocked O-star
wind to the entire X-ray emission from CSWs in \WR was only 1-2\% of
the observed X-ray flux.

In order to explore the wealth of physical processes that might be
important for the X-ray emission from CSWs in wide WR$+$O binaries as
\WRE, we ran a series of models that consider hot plasma in
collisional ionization equilibrium or with non-equilibrium ionization, 
and we also considered the case of partial heating of the electrons
at the shock front (non-equal electron and ion temperatures). We 
recall that the latter is adopted through the $\beta$-parameter that 
gives the ratio of the electron and mean plasma temperatures 
($\beta = T_e / T$) at the shock front and the electron and ion
temperatures equilibrate downstream behind the shock 
(see \citealt{zhsk_00} for details).

We fitted the X-ray spectra of \WR in \xspec with the CSW model
spectra using the nominal values of the wind parameters as described
above. In all the cases under consideration (different values of the
$\Lambda$-parameter), the theoretical spectra matched the shape of the 
observed spectra but they overestimated the observed flux by about two
orders of magnitude. In other words, the nominal wind parameters
suggest way too high an emission measure of the CSW region in \WRE.

We note that the emission measure is proportional to the square of the
plasma density (EM $\propto n^2 V$, $n$ is the number density, $V$ is
the volume), therefore, to the square of the stellar wind mass
loss. Alternatively, the emission measure is reversely proportional to
the binary separation ($n \propto 1/a^2$ and $V \propto a^3$,
therefore EM $\propto 1/a$) which suggests a much larger and
unrealistic binary separation for the \WR binary period of 4766 days
(Section \ref{sec:wrstar}) in order to synchronize the emission
measure with the observational requirements.
 
We thus re-ran the CSW spectral models by adopting the 
correspondingly reduced mass losses in order to match the observed
flux from \WRE. We also ran CSW spectral models with a reduced carbon
abundance to see its effect on the required mass-loss reduction
factor. Some fit results ($\Lambda = 36$) are given in 
Table~\ref{tab:csw} and  in Figure~\ref{fig:spec_csw}.

\begin{table*}
\caption{CSW Spectral Model Results 
\label{tab:csw}}
\begin{tabular}{lllllllll}
\hline
\multicolumn{1}{c}{} & 
\multicolumn{4}{c} {C / He $= 0.4$} & 
\multicolumn{4}{c} {C / He $= 0.1$} \\
\multicolumn{1}{c}{Parameter} & 
\multicolumn{1}{c}{CIE} & \multicolumn{1}{c}{NEI} &
\multicolumn{1}{l}{T$_{ei} +$CIE} & \multicolumn{1}{l}{T$_{ei} +$NEI} &
\multicolumn{1}{c}{CIE} & \multicolumn{1}{c}{NEI} &
\multicolumn{1}{l}{T$_{ei} +$CIE} & \multicolumn{1}{l}{T$_{ei} +$NEI} \\
\multicolumn{1}{c}{} &
\multicolumn{1}{c}{(A1)} & \multicolumn{1}{c}{(A2)} &
\multicolumn{1}{c}{(A3)} & \multicolumn{1}{c}{(A4)} &
\multicolumn{1}{c}{(B1)} & \multicolumn{1}{c}{(B2)} &
\multicolumn{1}{c}{(B3)} & \multicolumn{1}{c}{(B4)} \\
\hline
Reduced $\dot{M}$ by a factor of  &  12.7  & 12.7  & 9.5 & 9.5
                   &  9.5 & 9.5 & 7.7  & 7.7 \\
$\chi^2$/dof  & 190/165 & 176/165 & 143/165 &  142/165  
              & 190/165 & 176/165 & 157/165 &  158/165  \\
N$_{H}$ (10$^{21}$ cm$^{-2}$)  & 
          2.76$^{+0.23}_{-0.18}$ & 4.20$^{+0.20}_{-0.17}$ & 
          3.43$^{+0.20}_{-0.17}$ & 4.69$^{+0.20}_{-0.17}$ & 
          2.76$^{+0.27}_{-0.21}$ & 3.74$^{+0.23}_{-0.19}$ & 
          3.15$^{+0.23}_{-0.19}$ & 4.14$^{+0.23}_{-0.19}$ \\ 
O   & 0.16$^{+0.07}_{-0.05}$ & 0.13$^{+0.03}_{-0.03}$  
    & 0.08$^{+0.03}_{-0.03}$ & 0.09$^{+0.02}_{-0.02}$  
    & 0.05$^{+0.03}_{-0.02}$ & 0.04$^{+0.01}_{-0.01}$
    & 0.03$^{+0.01}_{-0.01}$ & 0.03$^{+0.01}_{-0.01}$
    \\ 
Ne  & 0.62$^{+0.13}_{-0.11}$ & 0.17$^{+0.07}_{-0.07}$  
    & 0.23$^{+0.05}_{-0.05}$ & 0.13$^{+0.04}_{-0.04}$  
    & 0.21$^{+0.05}_{-0.04}$ & 0.08$^{+0.03}_{-0.03}$
    & 0.10$^{+0.02}_{-0.02}$ & 0.06$^{+0.02}_{-0.02}$
    \\ 
Mg  & 1.17$^{+0.24}_{-0.20}$ & 0.47$^{+0.08}_{-0.08}$  
    & 0.28$^{+0.08}_{-0.07}$ & 0.19$^{+0.05}_{-0.04}$  
    & 0.43$^{+0.08}_{-0.07}$ & 0.22$^{+0.04}_{-0.03}$
    & 0.17$^{+0.04}_{-0.03}$ & 0.12$^{+0.02}_{-0.02}$
    \\ 
Si  & 1.81$^{+0.55}_{-0.48}$ & 1.10$^{+0.25}_{-0.23}$  
    & 0.41$^{+0.21}_{-0.19}$ & 0.39$^{+0.14}_{-0.13}$  
    & 0.69$^{+0.20}_{-0.15}$ & 0.50$^{+0.11}_{-0.10}$
    & 0.27$^{+0.09}_{-0.08}$ & 0.26$^{+0.07}_{-0.06}$
    \\ 
Fe  & 2.61$^{+0.56}_{-0.41}$ & 4.01$^{+0.60}_{-0.52}$  
    & 1.00$^{+0.19}_{-0.16}$ & 1.71$^{+0.27}_{-0.24}$  
    & 0.90$^{+0.21}_{-0.15}$ & 1.40$^{+0.21}_{-0.18}$
    & 0.46$^{+0.09}_{-0.07}$ & 0.76$^{+0.12}_{-0.10}$
    \\ 
$norm$ & 0.97$^{+0.09}_{-0.10}$ & 1.00$^{+0.08}_{-0.08}$ 
       & 0.99$^{+0.07}_{-0.07}$ & 0.99$^{+0.06}_{-0.06}$  
       & 1.09$^{+0.10}_{-0.11}$ & 1.06$^{+0.08}_{-0.08}$ 
       & 1.09$^{+0.08}_{-0.08}$ & 1.05$^{+0.07}_{-0.07}$ \\
F$_{X}$ ($10^{-13}$ ergs cm$^{-2}$ s$^{-1}$)  &
           \,\,2.92  & \,\,2.94  &
           \,\,2.66  & \,\,2.66  &
           \,\,2.90  & \,\,2.90  &
           \,\,2.69  & \,\,2.68  \\
                                              &
            (4.92) &  (7.57) &
            (5.60) &  (8.68) &
            (4.92) &  (6.52) &
            (5.23) &  (7.20) \\
\hline

\end{tabular}

Note --
Results from  simultaneous fits to the EPIC 
spectra of \WR using model spectra from the CSW hydrodynamic 
simulations. Two type of models were considered: one with the WC 
standard carbon abundance (C / He $= 0.4$) and another with a reduced
carbon abundance (C / He $= 0.1$). Four versions of each model were
adopted (A1 through A4 and B1 through B4): 
with collisional ionization equilibrium (CIE), 
with non-equilibrium ionization (NEI) as well as their versions that 
take into account the different electron and ion temperatures (T$_{ei}
+$CIE and T$_{ei} +$NEI). For each model, given is the factor by which 
the mass-loss rates of the stellar winds were reduced (see Section 
\ref{subsec:csw_spec} for details).
Tabulated quantities are the neutral hydrogen absorption column
density (N$_{H}$), the O, Ne, Mg, Si and Fe abundances, the 
normalization parameter ($norm$) and the absorbed X-ray flux (F$_X$) 
in the 0.5 - 10 keV range followed in parentheses by the unabsorbed 
value. The $norm$ parameter is a dimensionless
quantity that gives the ratio of observed to theoretical fluxes. A
value of $norm = 1.0$ indicates a perfect match between the observed
count rate and that predicted by the model.
The derived abundances are with
respect to the typical WC abundances \citep{vdh_86}.
Errors are the $1\sigma$ values from the fits.

\end{table*}

\begin{figure*}
\begin{center}
\centering\includegraphics[width=2.24in,
height=3.0in,angle=-90]{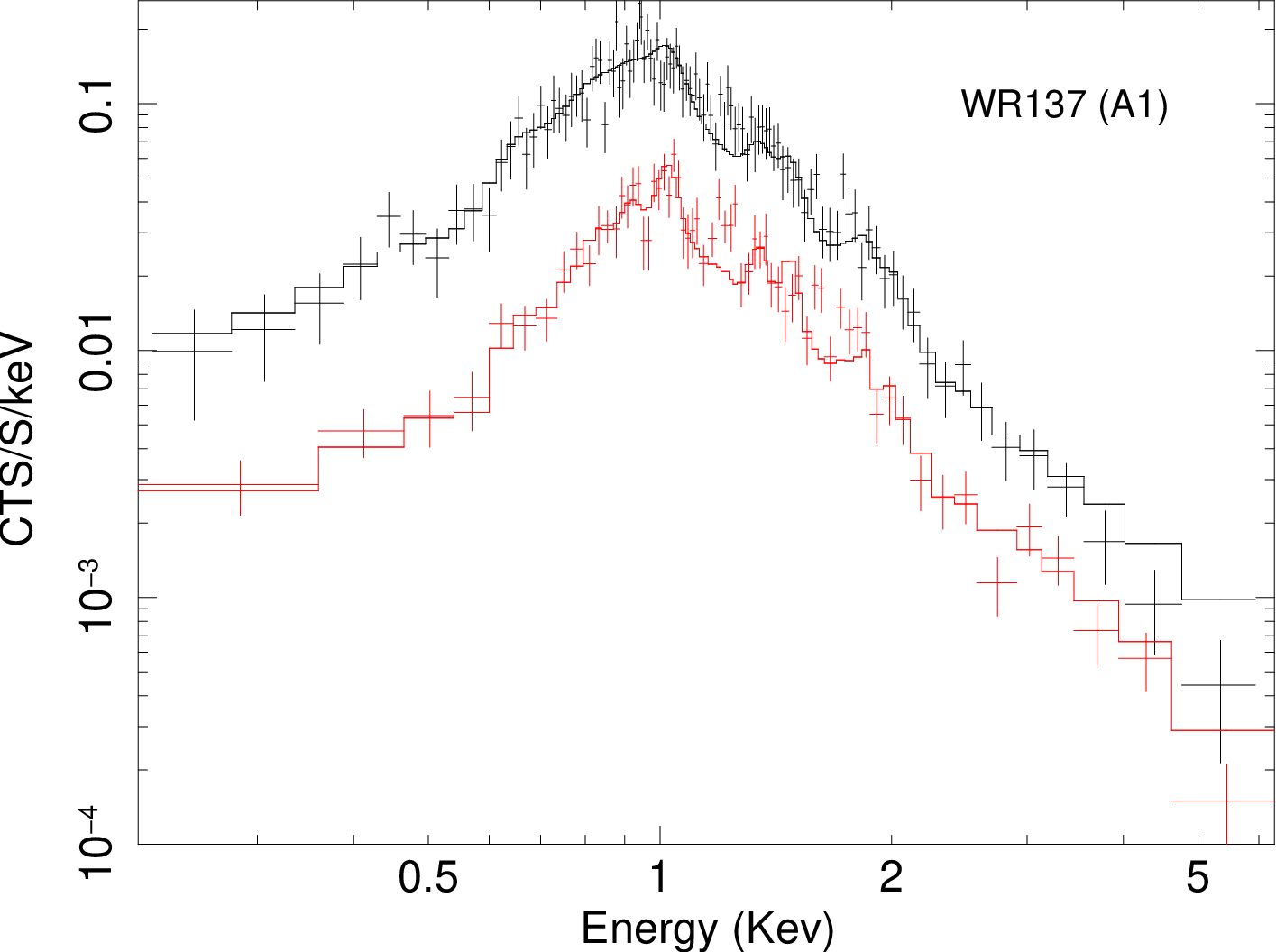}
\centering\includegraphics[width=2.24in,
height=3.0in,angle=-90]{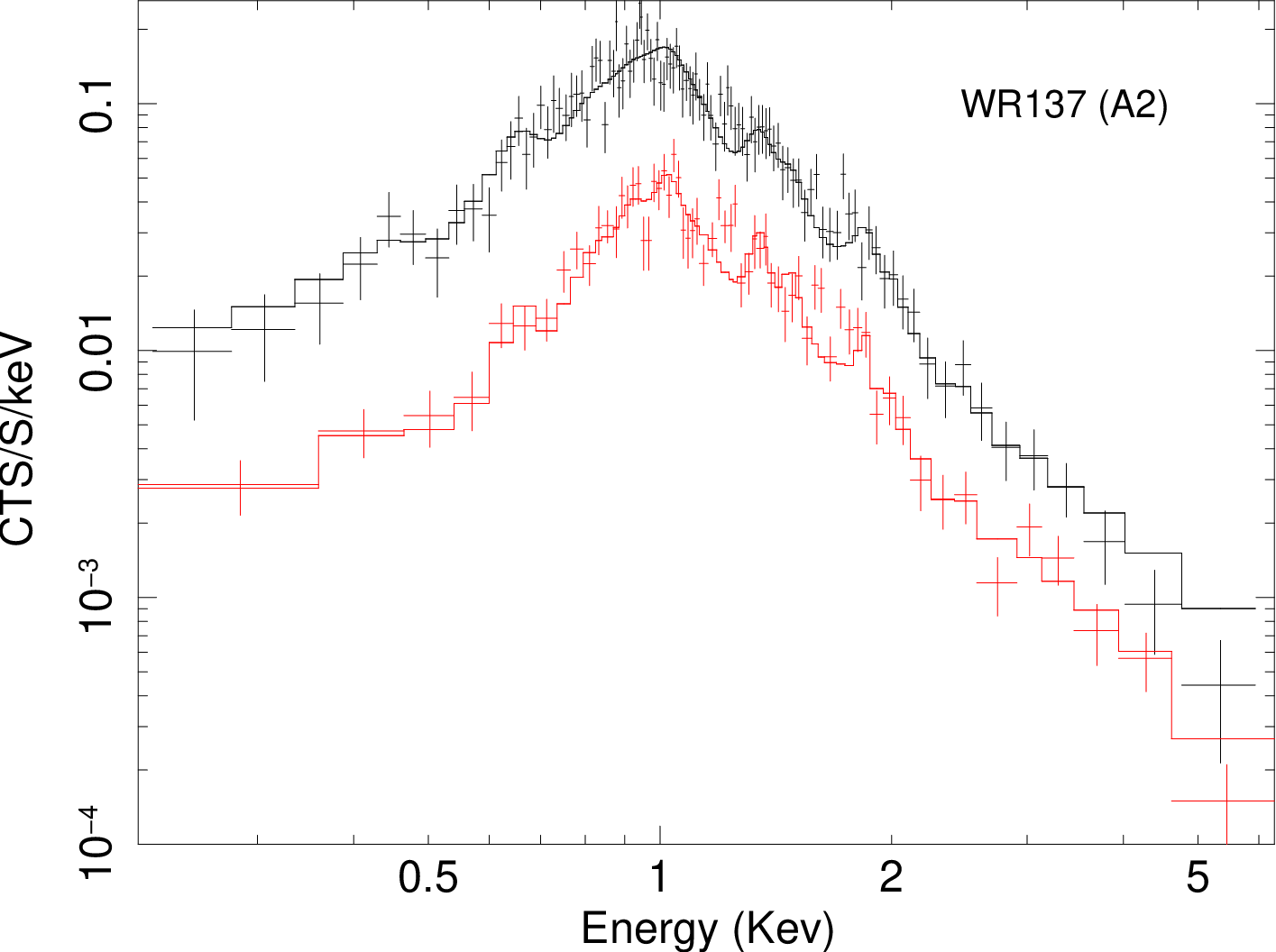}
\centering\includegraphics[width=2.24in,
height=3.0in,angle=-90]{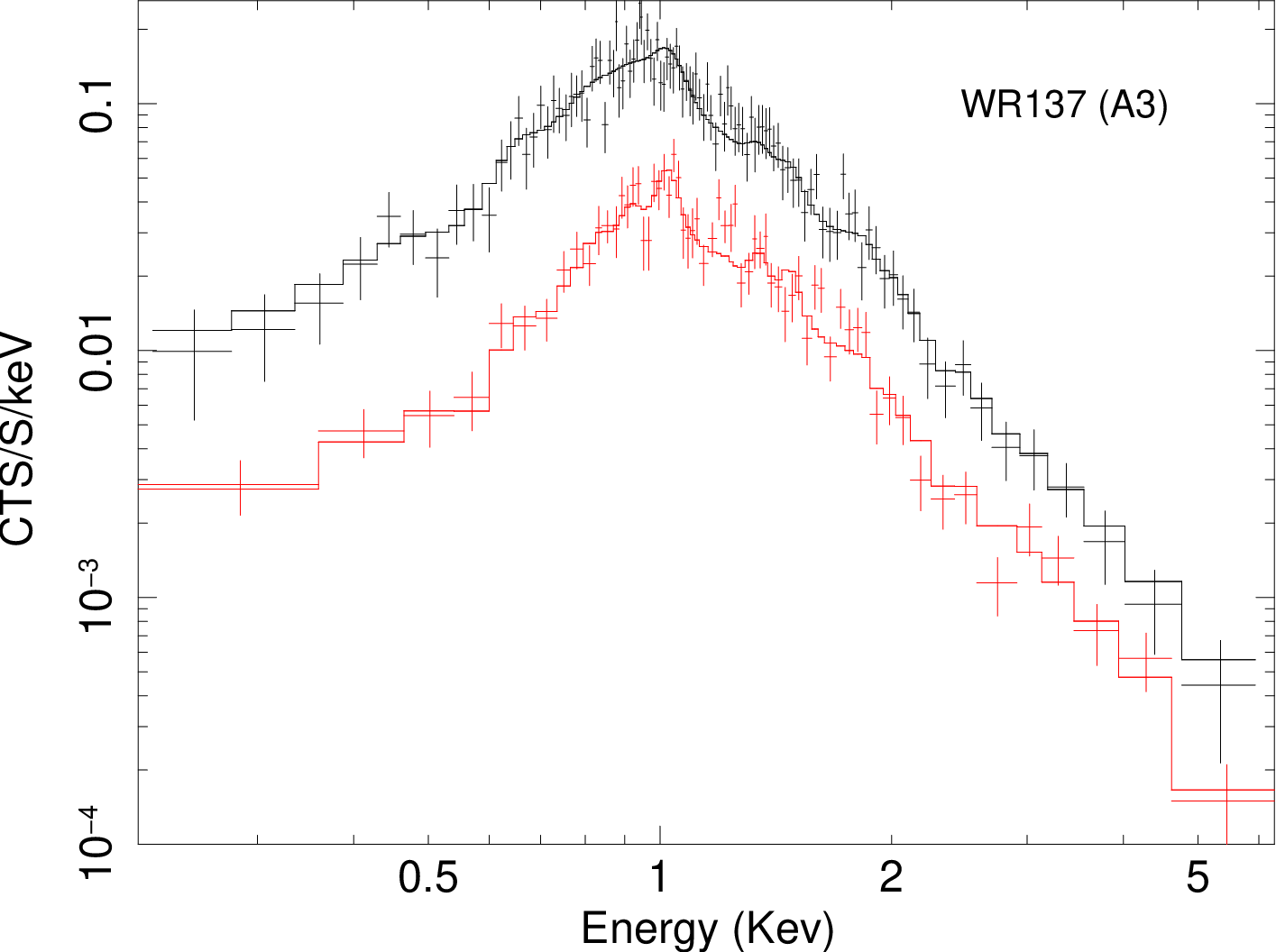}
\centering\includegraphics[width=2.24in,
height=3.0in,angle=-90]{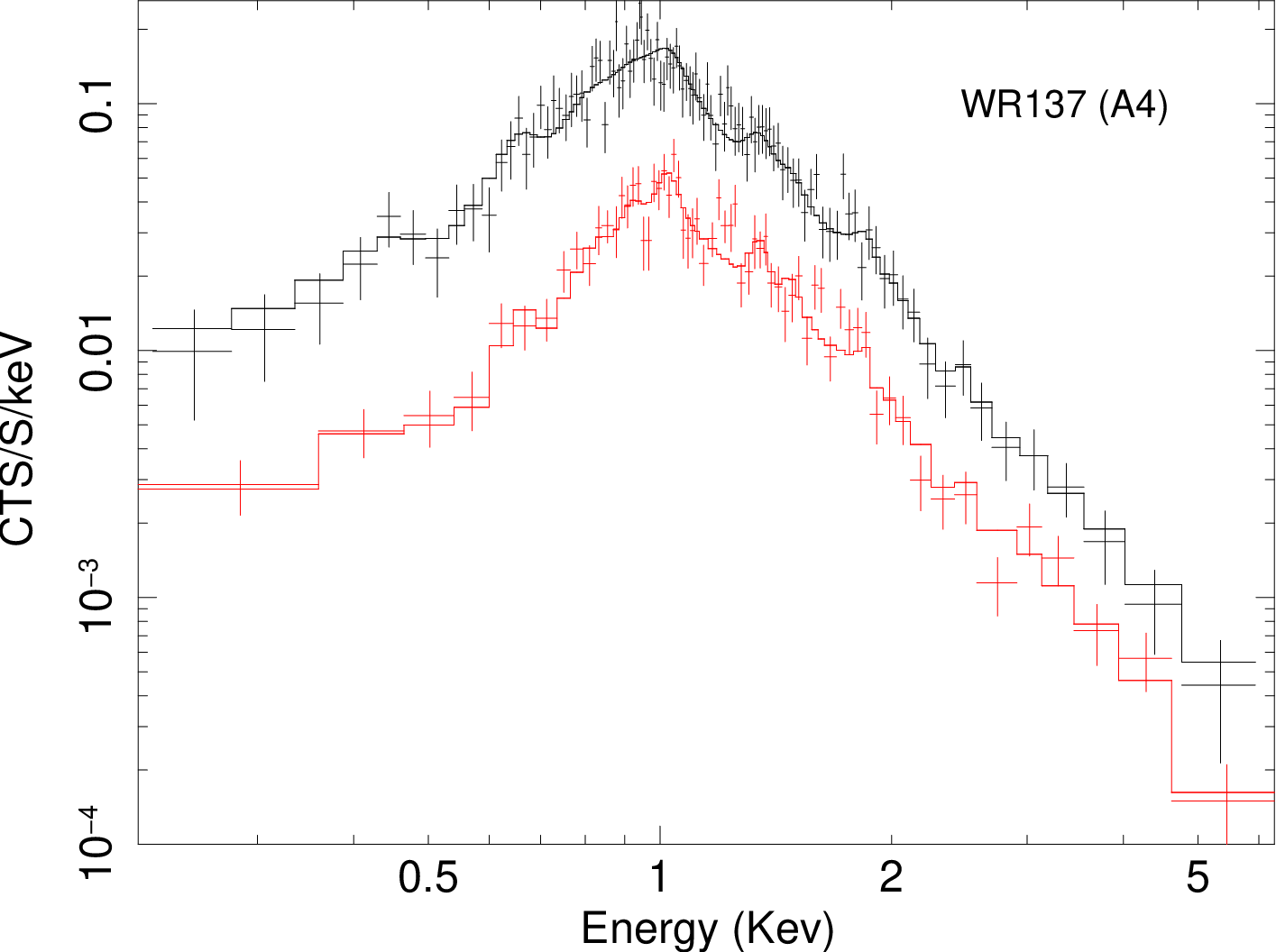}
\end{center}
\caption{The background-subtracted spectra of \WR  overlaid with the
CSW model fits.  The labels A1, A2, A3 and A4 denote the corresponding
models from Table~\ref{tab:csw}.
In each panel, the pn (upper curve) and MOS (lower curve) spectra are 
shown in black and red colour, respectively.
The spectra were re-binned to have a minimum of 30 counts per bin.
}
\label{fig:spec_csw}
\end{figure*}

We see that the CSW spectral models match acceptably well the observed
X-ray spectra of \WRE. This is valid for all the cases under
consideration: such with the carbon abundance typical for the WC stars 
or with the reduced one (respectively models A1-through-A4 and 
B1-through-B4 in Table~\ref{tab:csw}); such with CIE plasma or plasma
in NEI; such with plasma having different electron and ion temperatures.
However, the models with complete temperature equalization (models A1,
A2, B1 and B2) have the caveat of slightly overestimating the X-ray
emission from \WR at energies above $\sim 3$ keV (e.g.,
Fig.\ref{fig:spec_csw}). This is a likely indication of presence of
X-ray emission plasmas in the CSW region hotter than required by the 
\XMM spectra. Understandably, the models with a partial electron 
heating at the shock front (models A3, A4, B3 and B4
all having $\beta= T_e / T = 0.1$) provide a better match to the 
observed spectra. In fact, we ran CSW models with various values for
the $\beta$-parameter and the result was that all the models with 
$\beta \leq 0.1$ provided about the same quality of the CSW spectral
fits while those with higher degree of temperature equalization 
($\beta= 0.2 - 1$) started to show an excess emission at energies
above $\sim 3$ keV. 

For all the CSW model fits, the values of the X-ray absorption (N$_H$) 
are in general consistent with the optical extinction to \WR
(Section~\ref{sec:wrstar}). This is in fact in accord with the CSW 
picture in wide binaries in which the interaction region is not 
located deep in the stellar winds, thus, no appreciable wind 
absorption should be expected.

But, the most interesting result from adopting the CSW
picture in the case of \WR is probably the requirement of relatively
low mass loss in order to match the observed X-ray flux from this 
object. The mass loss needs be reduced by about one order of 
magnitude with respect to the currently accepted value for \WR
(Section~\ref{sec:wrstar}). Such a low figure ($\dot{M}_{WR} \approx
[2 - 4]\times10^{-6}$\dotM) is quite atypical for a WR star, so, could 
it indicate a very efficient wind clumping  even at large distance
from this WC star? That is, the stellar wind in \WR might be a
two-component flow and only the smooth (less massive) component is
playing a role for the colliding stellar winds in this binary.
On the other hand, the requirement for a very low mass loss might be 
a sign that the X-ray emission from \WR is due to some other mechanism 
rather than to CSWs in a wide WR$+$O binary.

Similar indication may come from the appreciable difference between 
the values of the X-ray luminosity of \WR as deduced from the \XMM 
data,
$L_X = 
4.5 \times10^{32}$ ergs s$^{-1}$, (the mean for the values in 
Table~\ref{tab:csw}) and from the previous observations: 
$L_X = 2.8\times10^{32}$ ergs s$^{-1}$ with {\it Einstein};
$L_X = 1.1\times10^{32}$ ergs s$^{-1}$ with {\it ROSAT} 
(the values cited in Section~\ref{sec:wrstar} were
rescaled to the distance of 2.38 kpc adopted in this study). 
This luminosity difference is hard to explain by the different
pass-bands of the X-ray telescopes. On the other hand,
it is worth noting that such a difference cannot be due solely to the
orbital changes since the binary separation was not too different in
these three occasions. Namely, the orbital phase and the corresponding
binary separation were 
$\phi = 0.338$ and $1.113 \times a$ (\XMME);
$\phi = 0.698$ and $1.083 \times a$ ({\it ROSAT}); 
$\phi = 0.775$ and $ 1.004 \times a$ ({\it Einstein}), where $a$ is 
the mean orbital separation. To find an explanation of this 
luminosity variability, we might need some physical 
reason (mechanism) for such changes that may go beyond the CSW 
picture in \WRE. However, we have to keep in mind that
the {\it Einstein} and {\it ROSAT}  data on \WR have very limited 
photon statistics. Future observations with good quality are indeed
needed to check this result which will help us improve our
understanding of the X-ray production mechanism in \WRE.

\subsection{Comparison with other EDMs}
\label{subsec:comparison}
Since \WR is one of the seven originally proposed episodic dust makers
\citep{williams_95}, it is interesting to compare its
global properties with those of other EDMs. We recall that the  global
properties of the prototype EDM, WR 140, in the infrared, radio and
X-ray spectral domains are best explained in the framework of the CSW
picture in a wide WR$+$O binary system \citep{williams_90}. 
With \WR having an X-ray spectrum with good quality, there are 
now three EDMs that were studied in those spectral domains 
in some detail: WR 48a, \WR and WR 140.

The IR emission is basic for EDMs: this is the spectral domain
where the classification of these objects came from 
\citep{williams_95}. The 
IR light curve of \WR \citep{williams_01} is quite similar to that of 
WR 48a \citep{williams_12}. Both LCs show a gradual increase 
and then decrease between the minimum and maximum values of their IR 
emission. They show some `mini eruptions' as well. In contrast, the 
IR emission of the prototype EDM, WR 140, surges to a maximum in a 
short period of time (due to the sudden onset of dust formation) and 
then decreases more slowly to its minimum value \citep{williams_90}.

The radio properties of these three EDMs are not uniform either. 
WR 140 is a strong and variable non-thermal radio source
(\citealt{williams_90}; \citealt{whi_be_95}; \citealt{do_05}). 
WR 48a is very likely a thermal (and variable) radio source
(\citealt{hindson_12}; \citealt{zh_etal_14}). 
\WR was first classified as a thermal radio source by
\citet{ab_86}. Later, \citet{do_00} listed it also as a non-thermal 
radio source, although this classification was based only on a private
communication. Thus, in detail radio studies of \WR are needed to
settle the origin of its radio emission.

On the other hand, the X-ray studies of EDMs are very important
since they provide information about the properties of the CSW region
in these presumable wide WR$+$O binaries. In X-rays, the most 
interesting feature is probably that the X-ray luminosity of \WR is 
appreciably lower than that of WR 48a and WR 140 (see
Section~\ref{sec:results}). While the orbital elements and the stellar
wind parameters of WR 48a are not well constrained \citep{zh_etal_14}, 
the prototype EDM, WR 140, is well studied in this respect. Thus, we 
can have a simple comparison between the X-ray luminosities of WR 140 
and \WR in the framework of the CSW picture. 

Namely, there exists a 
scaling law for the CSW X-ray luminosity with the mass-loss rate 
($\dot{M}$), wind velocity ($v$) and binary separation ($a$): 
$L_X \propto \dot{M}^2 v^{-3} a^{-1}$  (\citealt{luo_90}; 
\citealt{mzh_93}). We note that the stellar wind parameters and the
binary separations (or orbital periods) of \WR and WR 140 are not 
strikingly different (see Section~\ref{sec:wrstar} here and table 3
in \citealt{williams_90}). Based on this scaling law, it is then 
{\it not} 
anticipated that the X-ray luminosities of these objects will differ 
considerably (mean $L_X$[\WRE] / mean $L_X$[WR140] $\approx
0.89$).  
Nevertheless, we see that the X-ray luminosity of \WR 
is at least one order of magnitude less than that of WR 140: 
L$_X$ (\WRE) $\leq 10^{33}$ ergs s$^{-1}$ (Section~\ref{sec:results}) 
vs. 
L$_X$ (WR 140) $\geq 10^{34}$ ergs s$^{-1}$ (\citealt{zhsk_00};
\citealt{po_05}). 
In order to match the observed flux (and the X-ray luminosity) of
\WRE, an appreciable reduction of the mass-loss rate was needed in the
framework of the CSW picture. As mentioned in 
Section~\ref{subsec:csw_spec}, this raises questions and even doubts
about the overall validity of the CSW paradigm for the case of \WRE.
Along these lines, we mention that the X-ray spectrum of \WR could be well
represented by a two-component emission: one component is from thermal 
plasma and the second one is a `pure' continuum (black-body or
power-law) emission (see Section~\ref{sec:results}).

However, we have to keep in mind that the relatively strong X-ray
emission from \WR and the really strong one from WR 48a and WR 140 are
a clear sign that the binary nature of these objects provides some
suitable conditions for an efficient X-ray production mechanism to
operate in them. This
is so since these three objects (and all the EDMs as well) are 
carbon-rich Wolf-Rayet stars. And, we recall that all the pointed 
observations of presumably single WC stars resulted in non-detections 
(\citealt{os_03}; \citealt{sk_06}). Thus, single WCs are likely very 
faint or X-ray quite objects, opposite to what is observed from WR
48a, \WR and WR 140.

In summary, we note that although the CSW picture is very 
successful in explaining the global properties of the prototype 
EDM WR 140, we need to find some more observational support for its 
validity in such objects like \WR and WR 48a. We believe that studies 
of other EDMs, yet unobserved in X-rays, will be also very important 
to help us understand the physics of this extremely interesting 
phenomenon: episodic dust formation and its relation to colliding 
stellar winds in wide WR$+$O binaries.

\section{Conclusions}
\label{sec:conclusions}
In this work, we presented the \XMM data of \WR which provide the
first X-ray spectra with good quality of this episodic dust maker. The
basic results and conclusions are as follows.

(i) \WR is a relatively strong X-ray source 
($\log L_X = 32.76 - 32.83$, L$_X$ in ergs s$^{-1}$)
 with relatively weak emission-line features in its X-ray spectrum. 
It shows no short-term (within 35 ks) X-ray variability.

(ii) Discrete-temperature global spectral fits show that the X-ray 
spectra of \WR are well matched by a two-temperature optically-thin 
plasma emission (kT$_1 \sim 0.4$ keV and kT$_2 \sim 2.2$ keV). 
However, its X-ray emission is also well represented by a 
two-component model whose first component is from a thermal plasma 
and the second component is a `pure' continuum (black body or 
power-law) emission.

(iii) Colliding stellar wind model spectra match well the shape of the 
observed X-ray spectrum of \WRE. However, they overestimate the
observed flux (emission measure) by about two orders of magnitude. 
To reconcile the model predictions with observations,
the mass loss of \WR must be reduced considerably with respect to 
its currently accepted value. Such a low mass loss ($\dot{M}_{WR} 
\approx [2 - 4]\times10^{-6}$\dotM) may be indicates a very efficient 
wind clumping  even at large distance from this WC star.
Alternatively, it might be a sign that some other X-ray production
mechanism is at play in \WR rather than CSWs in a wide WR$+$O binary.

(iv) The X-ray absorption, derived from the spectral fits, is 
consistent with the optical extinction to \WRE. This is valid both for
the discrete-temperature and for the CSW spectral models.

(v) There are now three EDMs (WR 48a, \WR and WR 140) that have X-ray 
spectra with good quality. Since the X-ray properties of these 
objects are not quite uniform, X-ray studies of the entire group of 
EDMs will be very helpful for understanding the physics of episodic 
dust makers and to establish whether colliding stellar winds is their
basic X-ray production mechanism.

\section{Acknowledgements}
This research has made use of data and/or software provided by the
High Energy Astrophysics Science Archive Research Center (HEASARC),
which is a service of the Astrophysics Science Division at NASA/GSFC
and the High Energy Astrophysics Division of the Smithsonian
Astrophysical Observatory.
This research has made use of the NASA's Astrophysics Data System, and
the SIMBAD astronomical data base, operated by CDS at Strasbourg,
France.
The author thanks an anonymous referee for helpful comments and 
suggestions.

{}

\bsp

\label{lastpage}

\end{document}